\documentclass[reprint, twocolumn, prl, aps,showkeys,12]{revtex4-1}

\pdfoutput=1

\usepackage{graphicx}
\usepackage{graphics}
\usepackage{color}
\usepackage[scriptsize]{subfigure}
\usepackage{epstopdf}
\usepackage{ifthen}
\usepackage{todonotes}

\newcommand{\REM}[1]{\ifthenelse{0=1}{{#1}}{}}

\newcommand{\be}{\begin{equation}}
\newcommand{\ee}{\end{equation}}
\newcommand{\ba}{\begin{array}}
\newcommand{\ea}{\end{array}}
\newcommand{\bi}{\begin{itemize}}
\newcommand{\ei}{\end{itemize}}
\newcommand{\bea}{\begin{eqnarray}}
\newcommand{\eea}{\end{eqnarray}}

\newcommand{\br}{{\mathbf r}}
\newcommand{\QRtc}{$t_c=0.0760(5)$}
\newcommand{\QRz}{$z=1.99(5)$}
\newcommand{\QRnu}{$\nu=1.00(2)$}
\newcommand{\QReta}{$\eta=-0.3(1)$}
\newcommand{\QRw}{$w=0.6(2)$}
\newcommand{\QRwtwelve}{$w=0.92(7)$}

\newcommand{\QRyt}{$y_\tau=0.85(2)$}

\newcommand{\QRnumpoints}{142\ }
\newcommand{\QRnumMCS}{$6\times 10^4$ MCS\ }

\newcommand{\HCBhc}{$h_c=4.79(3)$}
\newcommand{\HCBz}{$z=1.88(8)$}
\newcommand{\HCBnu}{$\nu=0.99(3)$}
\newcommand{\HCBeta}{$\eta=-0.16(8)$}
\newcommand{\HCByr}{$y_r=1.718(1)$}

\graphicspath{{../Figures/}}

\begin{document}
\title{Quantum Critical Scaling of Dirty Bosons in Two Dimensions}
\author{Ray Ng}
\email{ngry@mcmaster.ca}

\author{Erik S. S{\o}rensen}
\affiliation{Department of Physics and Astronomy,
McMaster University
1280 Main Street West
L8S 4M1, Hamilton Ontario
CANADA}

\date{\today}

\begin{abstract}
We determine the dynamical critical exponent, $z$, appearing at the Bose glass
to superfluid transition in two dimensions by performing large scale numerical
studies of {\it two} microscopically different quantum models within the
universality class; The hard-core boson model
and the quantum rotor (soft core) model, both subject to strong
on-site disorder. By performing many simulations at different system size, $L$,
and inverse temperature, $\beta$, close to the quantum critical point, the
position of the critical point and the critical exponents, $z$, $\nu$ and
$\eta$ can be determined {\it independently} of any prior assumptions of the
numerical value of $z$. This is done by a careful scaling analysis close to the
critical point with a particular focus on the temperature dependence of the
scaling functions.  For the hard-core boson model we find \HCBz, \HCBnu\ and
\HCBeta\ with a critical field of \HCBhc, while for the quantum rotor model we
find \QRz, \QRnu\ and \QReta\ with a critical hopping parameter of \QRtc.  In
both cases do we find a correlation length exponent consistent with $\nu=1$,
saturating the bound $\nu\ge 2/d$ as well as a value of $z$ significantly
larger than previous studies, and for the quantum rotor model consistent with
$z=d$.

  \end{abstract}
\keywords{harcore-bosons, universality class}
\maketitle

Most familiar quantum critical points (QCP's) are characterized by Lorentz invariance
implying a symmetry between correlations in space and time and consequently
also between the respective correlation lengths $\xi\sim\xi_\tau$.  In turn, this
implies that the dynamical critical exponent, defined through $\xi_\tau\sim
\xi^z$, is simply $z=1$ although $\xi$ and $\xi_\tau$ typically differ by an
overall pre-factor. Systems for which $z\ne 1$ are comparatively less common
and possess a distinct anisotropy between space and time since $\xi$ and
$\xi_\tau$ are not only different but now also {\it scale} differently. 
One model for which it is generally believed that $z\ne 1$ is the Bose glass to
super fluid (BG-SF) transition describing interacting bosons subject to
disorder, the so called dirty-boson problem, modelled by the hamiltonian:
\begin{equation}
\label{eq:BH}
H_{bh}=-t\sum_{\br,\bf e} \left(b^{\dagger}_{\br}b_{\br+\bf e}+
h.c.\right)-\sum_{\br}\mu_{\br}\tilde n_{\br}
+\frac{U}{2}\sum_{\br}\tilde n_{\br}\left(\tilde n_{\br}-1\right).
\end{equation}
Here ${\bf e}={\bf x,y}$, and
$b^{\dagger}_{\br}, b_{\bf r}$ are the boson creation and annihilation operators at site $\br$
with $\tilde n_{\br}$ the corresponding number operator.
The parameters of the model are the hopping constant $t$, Hubbard repulsion
$U$, and {\it site}-dependent chemical potential $\mu_\br$, inducing the disorder.

Experimental setups
emulating dirty boson physics include optical lattices~\cite{Greiner02,*Pasienski10}
adsorbed
Helium in random media~\cite{Crowell97}, Josephson-junction arrays~\cite{Zant92},
thin-film superconductors~\cite{Haviland89,*Jaeger89,*Liu91,*Markovic98,*Lin2011} as well as quantum
magnets such as doped-DTN~\cite{Rongyu10,*Rong12nature,*Rong12}. 
For recent reviews see Ref.~\cite{Zheludev13,Weichman08}.

The dynamical critical exponent, $z$, appearing at the BG-SF transition has
proven exceedingly hard to determine. Initial theoretical
work~\cite{Fisher88,*Fisher90,*Fisher89} argued that $z=d$ in {\it any}
dimension. This has intriguing implications since it implies the absence of an
upper critical dimension, although a scenario involving a discontinuous onset of
mean field behavior has beeen
proposed~\cite{Dorogovtsev80,*Weichman89,*Mukhopadhyay96}. Subsequent numerical
studies~\cite{Krauth91,*Runge92,*Batrouni93,Sorensen92,*Wallin94,Zhang95,Kisker97,Prokofev04,Pollet09,*Gurarie09,*Carleo13,*Lin11}
of the system in two dimensions were consistent with $z=d=2$ and the phase-diagram is well understood~\cite{Bernardet02, *Sengupta07, *Soyler11}.
More recently, the transition
in three dimensions has been investigated both 
numerically~\cite{Hitchcock06,*Gurarie09,*Yao14} and experimentally~\cite{Demarco09},
yielding evidence
for $z=d=3$.
However, the
majority of these numerical studies were not unbiased since implicit
assumptions about $z$ are needed to fix the aspect ratio $L^z/\beta$ in the
simulations. 

The arguments leading to the equality $z=d$ starts with hyperscaling~\cite{Stauffer72,*Aharony74,*Hohenberg76} which
states that the singular part of the free energy inside a correlation volume is a 
{\it universal} dimensionless number, $(f_s/\hbar)\xi^d\xi_\tau=A.$ With $\xi\sim\delta^{-\nu}$
it follows that $f_s\sim\delta^{\nu(d+z)}$ with a finite-size form:~\cite{Fisher89}
\begin{equation}
f_s(\delta,L,\beta)\sim\delta^{\nu(d+z)}F(\xi/L,\xi_\tau/\beta).
\label{eq:fsscaling}
\end{equation}
Imposing a a phase gradient $\partial\phi$ along  one of the {\it spatial} directions will then
give rise to a free energy difference $\Delta f_s/\hbar = \frac{1}{2}\rho (\partial\phi)^2$ where
$\rho$ is the stiffness (superfluid density). Since $\Delta f_s$ must obey a form
similar to Eq.~(\ref{eq:fsscaling}) and since $\partial\phi$ has dimension of inverse length implying $\partial\phi\sim1/\xi$, 
it follows that $\rho\sim\xi^2\delta^{\nu(d+z)}\sim\delta^{\nu(d-2+z)}$, with a finite-size scaling form of:
\begin{equation}
\label{eq:rho}
\rho  = L^{d-2-z} R(\delta L^{1/\nu}, \beta/L^z),
\end{equation}
Following Ref.~\onlinecite{Fisher89},
an analoguous argument imposing a twist in the temporal direction leads to $\Delta f_s/\hbar=\frac{1}{2}\kappa (\partial_\tau\phi)^2$
and $\kappa\sim\xi_\tau^2\delta^{\nu(d+z)}\sim\delta^{\nu(d-z)}$ assuming $\partial_\tau\phi\sim 1/\xi_\tau$. On the other hand, if $\delta=(\mu-\mu_c)$, 
the {\it singular} part of the compressibility $\kappa_s$ must from Eq.~(\ref{eq:fsscaling}) obey $\kappa_s\sim\delta^{\nu(d+z)-2}$.
Assuming that $z>d$, so that $\kappa$
diverges, it follows that $\kappa=\kappa_s$ which leads to $z\nu=1$. With $z>d$ this result would then contradict
the (quantum) Harris inequality $\nu\ge 2/d$~\cite{chayes86} invalidating the initial assumption. 
Hence,  one must have $z\le d$. Finally, if one argues that for the disordered
system $\kappa$ cannot vanish at criticality the relation $\kappa\sim\delta^{\nu(d-z)}$ implies $z=d$.
\begin{figure}[t]
\includegraphics[width=\linewidth]{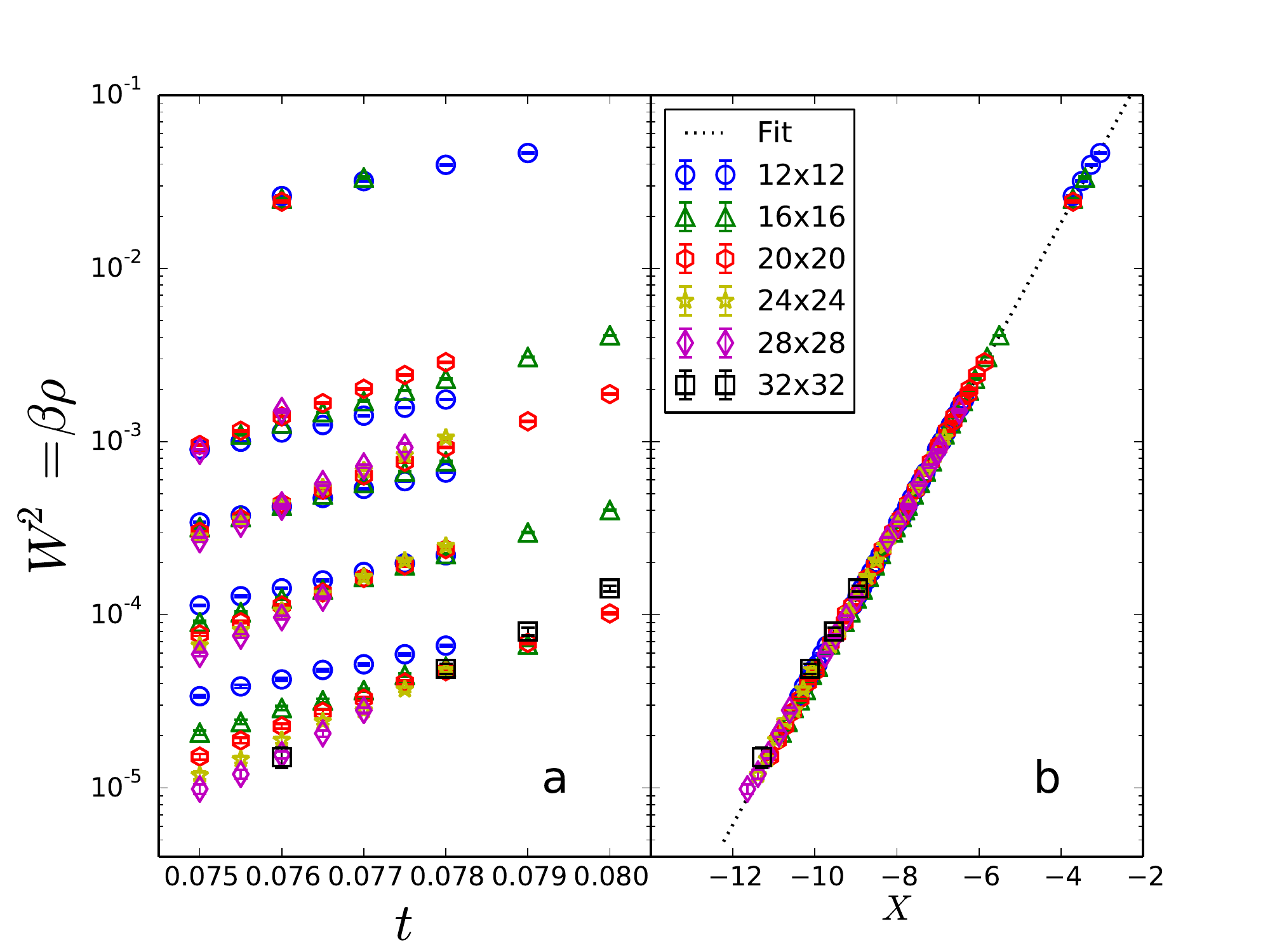}
\caption{
\label{fig:QRCollapse}
(Color online) Scaling collapse of \QRnumpoints\ independent simulations of $W^2=\beta\rho$ for the QR model. (a) Unscaled data
of $W^2$ versus $t$ for $L=12,\ldots,32$.
(b) Scaling collapse of the data of panel (a). The data is plotted
against the scaling variable $X=\ln(a\beta/L^z)+b(t-t_c)L^{1/\nu}-cL/\beta^{1/z}-d(L/\beta^{1/z})^2$ obtaining \QRz, \QRnu, \QRtc.
}
\end{figure}

More recent theoretical work~\cite{Weichman07, *Weichmanprb08,Weichman08} has
questioned the arguments leading to the equality $z=d$.  In particular, in the
presence of disorder breaking particle-hole symmetry, it was
argued~\cite{Weichman07, *Weichmanprb08} that $\kappa-\kappa_s$ is dominated by
the analytical background and $\partial_\tau\phi\sim 1/\xi_\tau$ should not
apply, invalidating the relation $\kappa\sim\delta^{\nu(d-z)}$, leaving $z$
unconstrained. A recent numerical~\cite{Priya06} study of the hard-core version
of Eq.~(\ref{eq:BH}) finds $z=1.4 \pm 0.05$, by analyzing scaling behavior of
quantities relatively far from criticality. However, in that study, relatively
few disorder realizations ($10-10^3$) were used and the location of the QCP was
not reliably determined. In more recent work, Meier et al~\cite{Meier12}
performed a state of the art calculation of a soft-core version of
Eq.~(\ref{eq:BH}) finding a significantly larger value of $z=1.75 \pm 0.05$ and $\nu=1.15(3)$.
While the location of the QCP was determined with an impressive precision this
latter study does not employ a fully quantum mechanical model but instead uses
an effective classical model for which the temperature dependence, at the heart
of the scaling with $z$, is only approximately accounted for.  It is for
instance not possible to calculate the specific heat of the underlying quantum
model using the representation of~\cite{Meier12}.

At present, the value of $z$ at the dirty-boson QCP along with many of the
other exponents most notably $\nu$ can therefore best be regarded as ill
determined, at least for the fully quantum mechanical model.  It is not known
to what extent, if any, the relation $z=d$ is violated nor if the relation
$\nu\ge 2/d$~\cite{chayes86} is obeyed, although, as outline above it seems
reasonable to expect $z\le d$ if indeed the inequality $\nu\ge 2/d$ is obeyed and, as shown in~\cite{Fisher89}, $z\ge 1$ in any dimension.
Here we try to answer these questions
by performing large-scale simulations on {\it two} fully quantum  mechanical
models; A hard-core boson model (HCB) modelled as a transverse field $XY$ model
and a soft-core quantum rotor model (QR), both of which are subject to strong
on-site disorder.  In all cases do we use $10^4-10^5$ disorder realizations
over a large range of temperatures extending down to $\beta=1024$ for linear
system sizes $L=12-32$.

\begin{figure}[t]
\includegraphics[width=\linewidth]{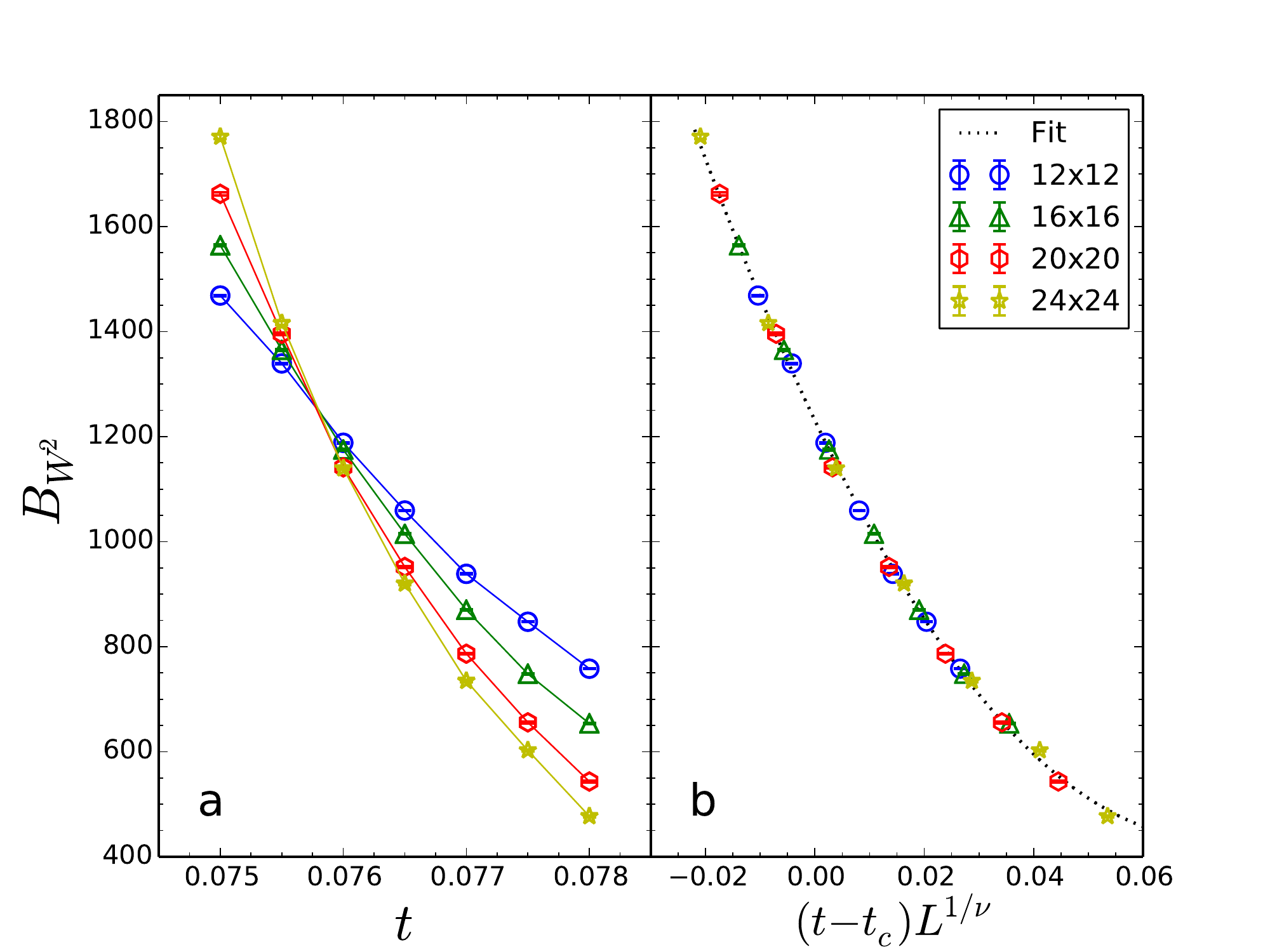}
\caption{
\label{fig:QRBinder}
(Color online) The Binder cumulant 
$B_{W^2}$ for the QR model versus $t$ with
$\beta = L^2/4$. (a) Unscaled data showing a crossing
close to the critical point \QRtc. (b) Scaling plot versus $(t-t_c)L^{1/\nu}$ obtained by fitting the data in (a) to the
form $a+b(t-t_c)L^{1/\nu}+c(t-t_c)^2L^{2/\nu}$ yielding $t_c=0.758(5)$ and $\nu=0.98(3)$.
}
\end{figure}


{\it Models:} 
The first model we study, closely related to Eq.~(\ref{eq:BH}), is the quantum
rotor (QR) model. It is defined in terms of conjugate phase and number
operators $\theta_\br,n_\br$ satisfying
$[\theta_\br,n_{\br'}]=\delta_{\br,{\br'}}$ on a $L\times L$ lattice:
\begin{equation}
H_{\text{qr}}=
-\sum_{{\br},{\br+\bf e}} t \cos(\theta_{\bf r}-\theta_{\bf r+e})
-\sum_{\bf r}\mu_{\bf r}n_\br
+\frac{U}{2}\sum_{\bf r} n^2_\br \,
\label{eq:HQR}
\end{equation}
where $U$ is the onsite repulsion, $t$ is the nearest neighbor tunneling
amplitude and $\mu_r\in[-\Delta, \Delta]$ represents the uniformly distributed
on-site disorder in the chemical potential. As before, ${\bf e}={\bf x,y}$.
The disorder for a given disorder realization is not constrained and in all
simulations we use $\Delta=\frac{1}{2}$, $U=1$ and tune through the BG-SF
transtion varying $t$ at constant $\Delta$.  In contrast to Eq.~(\ref{eq:BH}) $n_\br$ can take
negative as well as positive values and one can loosely associtate $n_\br$ with
deviations from the average filling $n_0$ in Eq.~(\ref{eq:BH}), $\tilde
n_\br-n_0$.  For convenience we study Eq.~(\ref{eq:HQR}) using a link-current
representation~\cite{Krempa14} for which {\it directed} worm algorithms are
available~\cite{Aleta,*Aletb}. We use lattice ranging from $L=12$ to $L=32$,
with 50,000 disorder realizations for $L=12,\ldots, 28$ and 10,000 disorder
realizations for $L=32$ in all cases with \QRnumMCS\ per disorder realization.
For the simulations of the QR model a temporal discretization of $\Delta\tau=0.1$ was
used, sufficiently small that remaining discretization errors could be neglected.

The second model we consider is the $U\to\infty$ hard-core limit of
Eq.~(\ref{eq:BH}) where the boson occupation number is constrained to $0,1$. It
is therefore therefore equivalent to the following $S=1/2$ $XY$-model on a $L\times L$ lattice in a
random transverse field:
\begin{equation}
H_{xy}=-\frac{1}{2}\sum_{\br,\bf e} \left(S^{+}_{\br}S^{-}_{\br+\bf e}+S^{-}_{\br}S^{+}_{\br+\bf e}\right)+\sum_\br h_\br S^z_\br,
\end{equation}
where $h_\br\in \left[-h,h\right]$ uniformly. In this case we tune through the transition
by increasing the disorder strength, $h$. Despite the representation as a
spin model, we shall refer to this as the hard-core boson model (HCB).  We use
a directed loop version of the stochastic series expansion
(SSE)~\cite{Sandvik91,*Olav02} to simulate this model. This technique does not have
discretization errors and efficient directed algorithms~\cite{Olav02,Alet05} are available. For the SSE
calculations, we use a beta-doubling scheme~\cite{Sandvik02} that allows us to
very quickly equilibrate at large $\beta$ values. 
For each temperature in the beta-doubling scheme, we average over $48$ Monte Carlo sweeps (MCS),
with each sweep consisting of one diagonal update and $N_l$ directed loop
updates. $N_l$ is set during the equilibration phase so that on average $2
\langle n_H \rangle$ vertices are visited, where $n_H$ is the number of
non-trivial operators in the SSE string. 
In contrast to the QR model, we use here a microcanonical ensemble for the disorder by constraining each
disorder realization to have {\it exactly} $\sum h_\br =0$. This facilitates
the analysis and is not believed to affect the
results~\cite{Dhar03,*Monthus04}. 
We use at least $\sim 10^5$ disorder realizations per data point, a large improvement over~\cite{Priya06}.

In the following $[\ldots]$ denotes the disorder average while $\left<\ldots\right>$ denotes the thermal
average. In simulations of both models two independent replicas $\alpha,\beta$ of each disorder realization are simulated in parallel
so that combined averages $[\left<\ldots\right>^2]$ may be correctly estimated as $[\left<\ldots\right>_\alpha\left<\ldots\right>_\beta]$.

\begin{figure}[t]
\includegraphics[width=\linewidth]{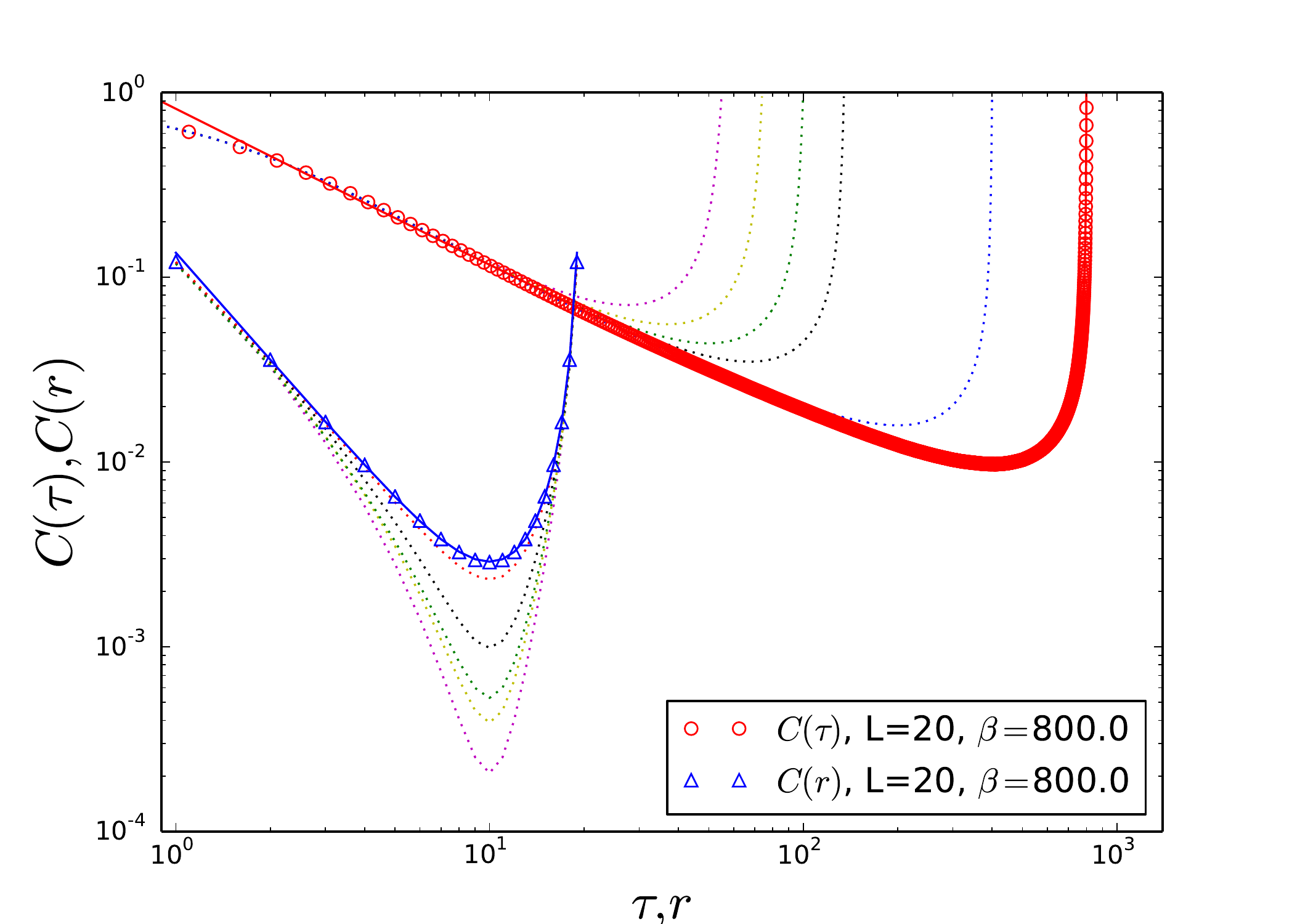}
\caption{
\label{fig:QRCt}
(color online) The correlation functions $C(\tau)$ and $C({\bf r})$ as a function of $\tau,{\bf r}$ for a system size $L=20$. Results 
are shown for the QR model at the critical
point and a range of $\beta=55,\ldots,800$. 
The solid red line is a  fit to $\beta=800$ results for $C(\tau)$ using the form $a(\tau^{-y_\tau}+(\beta-\tau)^{-y_\tau})$ with
$(z+\eta)/z$=\QRyt.
}
\end{figure}


{\it Observables:}
Our main focus is the scaling behaviour of the superfluid stiffness, $\rho$, for which
the finite-size scaling form Eq.~(\ref{eq:rho}) was derived. For both models we measure
$\rho$ as:
\begin{equation}
\rho=\frac{[\left<W_x^2+W_y^2\right>]}{2\beta},
\label{eq:rw}
\end{equation}
where $W_x$ and $W_y$ are the winding numbers in the spatial directions. (For the HCB model Eq.~(\ref{eq:rw}) is
multiplied by $\pi$ to yield $\rho$.) From Eq~(\ref{eq:rw}) it follows that $\beta\rho=W^2$ has a
particularly attractive scaling form when $d=2$, which we may write:
\begin{equation}
\label{eq: rho}
W^2  = \frac{\beta}{L^z} W(\delta L^{1/\nu}, L/\beta^{1/z}),
\label{eq:W2}
\end{equation}
where we define $\delta=(t-t_c)$ (QR model) and $\delta=(h-h_c)$ (HCB model). 
We also make extensive use of the correlation functions, defined as $C(\br-\br',\tau-\tau')=[\left<\exp(i(\theta_\br(\tau)-\theta_{\br'}(\tau')))\right>]$
for the QR model and as $C(\br-\br',\tau-\tau')=[\left<S^+_\br(\tau),S^-_{\br'}(\tau')\right>]$ for the HCB model.


{ \it \bf{Results, QR:}}
A large number of independent simulations of Eq.~(\ref{eq:HQR}) were carried
out at different $L=12,\ldots,32$ and $\beta=20,\ldots,400$ close to the QCP.
Since we expect $\rho$ to approach zero in an exponential manner as $L$ is
increased at {\it fixed} $\beta$ and since $\rho$ is likely exponentially
suppressed in the insulating phase it seems reasonable to approximate the
function $W(x,y)$ in Eq.~(\ref{eq:W2}) as $a\exp(f(x,y))$ with $x=\delta
L^{1/\nu},\ y=L/\beta^{1/z}$.  If the temperature dependence is carefully
mapped out~\cite{Supp} one indeed sees that $W(x,y)$ has a clear exponential
dependence.  As a first step, we then assume $f(x,y)=b x - c y- d y^2$. We can
then fit {\it all} \QRnumpoints\ data points to this form determining the
coefficients $a,b,c,d$ along with \QRtc\, \QRnu\ and \QRz. The results are
shown in Fig.~\ref{fig:QRCollapse} with a scaling plot using the scaling
variable
$X=\ln(a\beta/L^z)+b(t-t_c)L^{1/\nu}-cL/\beta^{1/z}-d(L/\beta^{1/z})^2$. A more
refined analysis~\cite{Supp} shows that the temperature dependence likely involves a
correction term $W^2=a y^z\exp(b x-c y)+ d y^{-w}\exp(b x - c' y)$. The
correction term is here proportional to $T^w$ and disappears as $T$ tends to
zero. It is straight forward to fit {\it all} our data to this form which yields
identical estimates for $t_c,\nu$, $z$ along with \QRw. Estimating the AIC (Akaike Information Criterion)
for the two forms heavily favors the latter.

With a reliable estimate of $z$ we can {\it fix} the scaling argument
$L^z/\beta$ by appropriately selecting $\beta$ for each $L$. If we then study
the Binder cumulant $B_{W^2}=[\left<W^4\right>]/[\left<W^2\right>]^2$ we see
that at {\it fixed} $L^z/\beta$ it should follow a simplified form of
Eq.~(\ref{eq:W2}), $B_{W^2}=B(\delta L^{1/\nu})$. As shown in
Fig.~\ref{fig:QRBinder}, lines for different $L$ will then cross at $t_c$.
This is indeed the case, confirming our previous estimates.

Our results for the correlation functions for the QR models are shown in
Fig.~\ref{fig:QRCt} for a $L=20$ lattice at $t_c$ for a range of temperatures.
Asymptotically, one expects~\cite{Fisher89} $C(\tau)\sim\tau^{-(d+2-z+\eta)/z}$
and $C(\br)\sim r^{-(d+2-z+\eta)}$.  Clearly, $C(\br)$ drops off much faster than
$C(\tau)$ confirming that $z\ne 1$. However, pronounced finite temperature
effects are visible in $C(\br)$ arising because the limit $\beta\gg L^z$ has not
yet been reached and we have not been able to reliable determine the power-law
describing $C(\br)$.  However, from $C(\tau)$ we determine $(z+\eta)/z$=\QRyt\ and
hence \QReta\ using our previous estimate \QRz.

For the QR model we have also verified that the compressibility, $\kappa$,
remains finite and independent of $L$ throughout the transition, consistent
with $z\le d$. Furthermore, a direct evaluation of $\frac{\partial W^2}{\partial t}$
directly at $t_c$ for fixed $L^z/\beta$, expected from Eq.~(\ref{eq:W2}) to scale as $\sim L^{1/\nu}$,
yields $\nu=0.98(4)$ consistent with our previous results. 

\begin{figure}
\includegraphics[width=\linewidth]{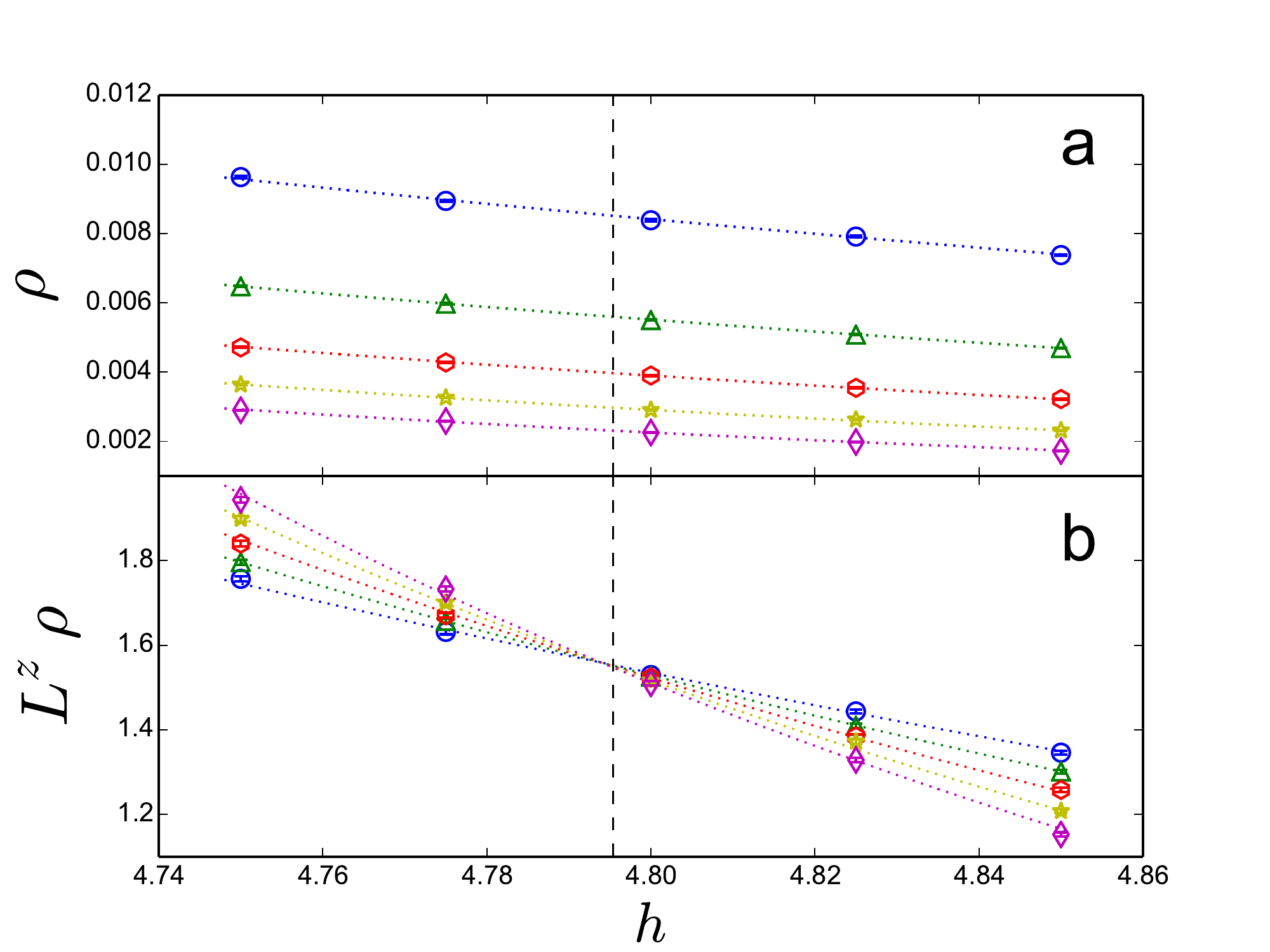}
\caption{(Color online) Finite-size scaling analysis of the stiffness for the
HCB model using data only at $\beta=512$. (a) $\rho$ vs $h$ for system sizes:
$L=16,20, \dots 32$. (b) Crossing of the scaling function: $\tilde{R}=L^z
\rho$ at the critical point: \HCBhc\ (vertical dashed line).
For both panels, the dotted lines are second order series expansions of
$\tilde{R}(x)= a + b x +c x^2$,
fitted to the data.
}
\label{fig:HCBstiff}
\end{figure}

{ \it \bf{Results, HCB:}}
Due to the hard-core constraint number fluctuations are dramatically suppressed in the HCB model. 
Combined with the very effective beta-doubling scheme we can
effectively reach much lower temperatures with the HCB model relative to the QR model.
Hence, we use a simplified form of Eq.~(\ref{eq:rho}):
\begin{equation}
\label{eq:hcbrho}
\rho =L^{d-2-z} \tilde R(\delta L^{1/\nu}),
\end{equation}
suppressing the temperature dependence. We have extensively verified that this is permissible for
the system sizes used~\cite{Supp} and that our data appear independent of temperature at $\beta=512$ to within
numerical precision.
We then fit our data for $\rho$ at $\beta=512$ to an expansion of $\tilde R$ in Eq.~(\ref{eq:hcbrho}) to second order: $\tilde R = a + b \delta L^{1/\nu}
+ c (\delta L^{1/\nu})^2$ obtaining the estimates: \HCBhc, \HCBz, \HCBnu. The
result of this fit is shown in Fig.~\ref{fig:HCBstiff}, where our simulation
data (unfilled markers) is superimposed with the quadratic fit
(dotted lines). In panel $b$ of Fig.~\ref{fig:HCBstiff}, we show the crossing
of the scaling function $\tilde{\rho}$, at \HCBhc. By including all results for $\beta<512$ it is also possible to perform an identical analysis
to the one performed for the QR model in Fig.~\ref{fig:QRCollapse}. Such an analysis similar results for $z$, $\nu$ and $h_c$.

\begin{figure}
\includegraphics[width=\linewidth]{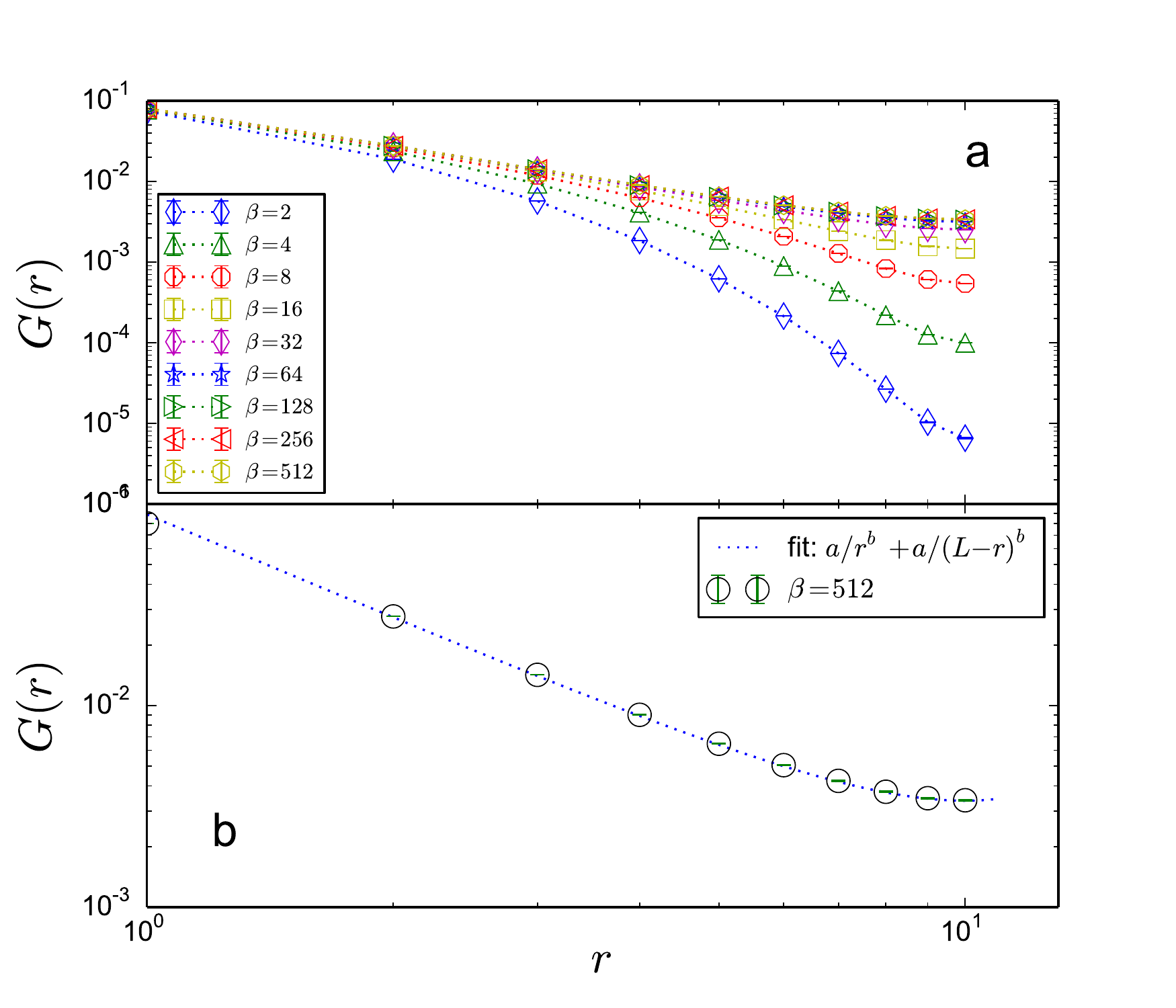}
\caption{(color online). The equal-time spatial correlation function: $C(\br)$ of
a $20 \times 20$ lattice for the HCB model at $h_c$. (a) Convergence of $C(\br)$ using the beta-doubling
procedure. For $L=20$, $C(\br)$ appears independent of $\beta$ for $\beta\ge 128$. (b) $C(\br)$
simulation data (open circles) at $\beta=512$ fitted to the form $a\left(1/r^{y_r}
+ 1/(L-r)^{y_r }\right)$ (dashed line), yielding $z+\eta=$\HCByr.}
\label{fig:HCBcorr}
\end{figure}

As was the case for the QR model the correlation functions show a pronounced
temperature dependence as shown in Fig.~\ref{fig:HCBcorr}({\bf a}) for $C(\br)$. However,
as we lower the temperature, $C(\br)$ reaches a stable power-law form at $\beta=512$ for {\it all} $L$ studied
showing that the regime $\beta\gg L^z$ is reached for all $L$ and
confirming that the temperature dependence can be neglected in Eq.~(\ref{eq:hcbrho}).
To determine the anomalous dimension $\eta$ we then fit the results in Fig.~\ref{fig:HCBcorr}({\bf a}) for $L=20$, $\beta=512$
\HCBhc\ to a power-law form with $z+\eta$=\HCByr\ as shown in Fig.~\ref{fig:HCBcorr}({\bf b}). 
Using our earlier estimate of $z$, we obtain \HCBeta\ in reasonable agreement with the result for the QR model.

For the HCB model we have also calculated the compressibility, $\kappa$. It remains roughly constant and independent of $L$ through
the transition.

{\it Conclusion:}
Our results for $\nu$ for both models studied indicate clearly that $\nu\ge 2/d$ is satisfied as an {\it equality}.
For the dynamical critical exponent $z$, describing the BG-SF transition, we find a value that is significantly larger than previous estimates.
While there is a slight disagreement in the estimate of $z$ for the two models we studied it seems 
possible that indeed $z=d$. 
In light of this it now seems particularly interesting to focus attention on the transition in $d=4$.


During the final stages of writing this manuscript we became aware of Ref.~\onlinecite{Alvarez14} which for the HCB model reach conclusions similar to ours.

\section{Acknowledgements}
\acknowledgments{We would like to thank Fabien Alet and Rong Yu for useful discussions.
This work was supported by NSERC and made possible by the facilities of the
Shared Hierarchical Academic Research Computing Network
(SHARCNET:www.sharcnet.ca) and Compute/Calcul Canada.}


\bibliography{refs}

\appendix
%
\newpage
\section{Supplementary material}
\label{sn:supplementary material}

\subsection{HCB simulation details}

\begin{figure}
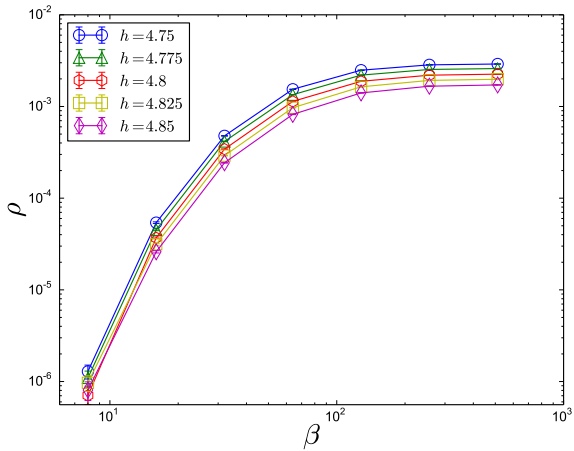

\includegraphics[width=\linewidth]{{{32x32nH_betadoubling-replica-h6}}}
\caption{Demonstration of the convergence of the spin stiffness: $\rho$ versus
$\beta$ using the beta-doubling procedure, for a $32\times32$ lattice.  As
$\beta$ increases, thermal effects can be clearly seen to disappear,  thereby
justifying the use of the simplified scaling form in Eq.~(\ref{eq:hcbrho}).}
\label{fig:HCBstiffbetadoubling}
\end{figure}

A central assumption made in our analysis of the HCB results was that a simplified scaling
function, Eq.~(\ref{eq:hcbrho}),  for $\rho$. In the main text, we showed the
convergence of the equal-time spatial correlation function: $C(\br)$ in
Fig.~\ref{fig:HCBcorr} at the critical point using data from the beta-doubling
procedure. For completeness, we show the convergence of the stiffness in
Fig.~\ref{fig:HCBstiffbetadoubling} for a $ 32\times 32$ lattice. Unlike
the $20\times 20$ system, we need to go up to $\beta=512$ before finite
temperature effects are eliminated. The data indicates that for the simulation
of even larger system sizes, one may need to go up to $\beta=1024$ to validate
the use of the simplified scaling form.

\begin{figure}
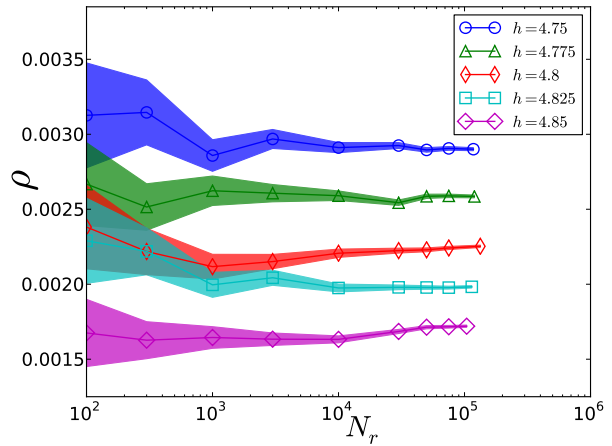

\includegraphics[width=\linewidth]{{{convergenceL32h}}}
\caption{Convergence of spin stiffness: $\rho$ with number of disorder
realizations, $N_r$, for $h$ values used. The shaded region indicates the
errorbars of points which decrease as more samples are used. Note that for the
equilibration step size chosen in our simulations, $\sim 10^5$ disorder
realizations are necessary to obtain true convergence. 
}
\label{fig:convergence}

\end{figure}

Another important point that is often overlooked, is the convergence of
observable data with the number of disorder realizations, $N_r$ used. For the
SSE parameters chosen, we find that it was necessary to use at least $\sim 5
\times 10^4$ disorder samples before disorder fluctuations are reasonably
small. This is demonstrated in Fig.~\ref{fig:convergence} for the largest
lattice size: $32\times 32$. For reliable data with controlled errorbars, all
the HCB data points were averaged over least $10^5$ independent disorder
realizations. This is contrast to earlier studies~\cite{Priya06} most relevant
to our work, where only $10^2-10^3$ disorder realizations were used. 
We note that it is quite unlikely that self-averaging applies in this model
and increasing the lattice size does therefore not decrease the number of
disorder realizations needed.

\section{Temperature dependence of $W^2=\beta\rho$}
\begin{figure}
\includegraphics[width=\linewidth]{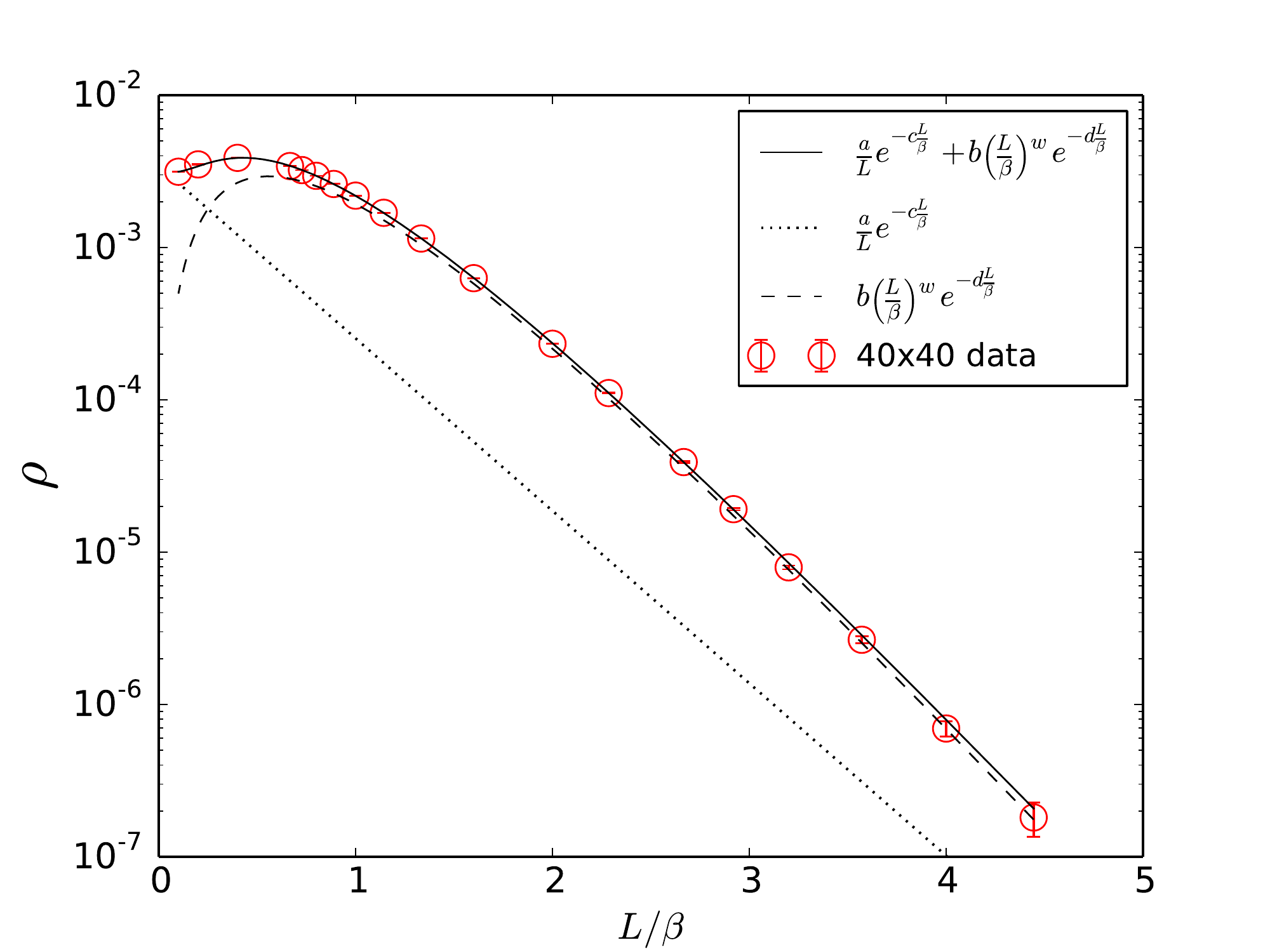}
\caption{
\label{fig:QRLStiff40}
(Color online) $\rho$ as a function of $L/{\beta}$ for the QR model {\it without} disorder. 
Results are shown for $L=40$.
The solid black line indicates a fit to the $L=40$ data of the form $\frac{a}{L} e^{-c \frac{L}{\beta}}  + b \left(\frac{L}{\beta}\right)^{y} e^{-d\frac{L}{\beta}}$
yielding $y=1.97(9)$.
The dotted line indicates the first part of this fit while the dashed line
shows the second part.
}
\end{figure}

Insight into the temperature dependence of $W^2$ can be gained by first studying
$\rho$ for the QR model {\it without} disorder. A model for which it is know that $z=1$.
Results for $\rho$ for a $40\times 40$ lattice are shown in Fig.~\ref{fig:QRLStiff40} for $\beta=9,\ldots 400$.
For $\beta\ll L$ we expect $\rho$ to go zero in an exponential manner while for $\beta \gg L$ it should
approach a constant. The simplest ansatz is therefore:
\begin{equation}
\rho= \frac{a}{L} e^{-c\frac{L}{\beta}},
\end{equation}
where we have tentatively included an $L$ dependence.
However, as is clearly evident in  Fig.~\ref{fig:QRLStiff40},
$\rho$ has a {\it maximum} close to $L/\beta=0.4$. The presence of this maximum signals
that there are likely two contributions to $\rho$ describing the $\beta \ll L$ and $\beta \gg L$ regimes.
(Although we note that the existence of two terms does not imply a maximum.)
We therefore assume the presence of a correction term proportional
to $T^y$ with $y>0$. Such a term will therefore disappear in the zero temperature limit. For
our final ansatz we therefore take:
\begin{equation}
\rho= \frac{a}{L} e^{-c\frac{L}{\beta}}+b\left(\frac{L}{\beta}\right)^y e^{-d\frac{L}{\beta}}.
\end{equation}
This is the form used in Fig.~\ref{fig:QRLStiff40} and it  gives an essentially {\it perfect} fit over the entire range of the figure.
We can immediately generalize to a scaling form for $W^2=\beta\rho$:
\begin{equation}
W^2= a\frac{\beta}{L} e^{-c\frac{L}{\beta}}+b\left(\frac{L^z}{\beta}\right)^w e^{-d\frac{L}{\beta}}.
\end{equation}
A fit to this form yields $\omega=0.97(9)$ in relative good agreement with similar correction terms used in Ref.~\cite{Krempa14}.

\begin{figure}
\includegraphics[width=\linewidth]{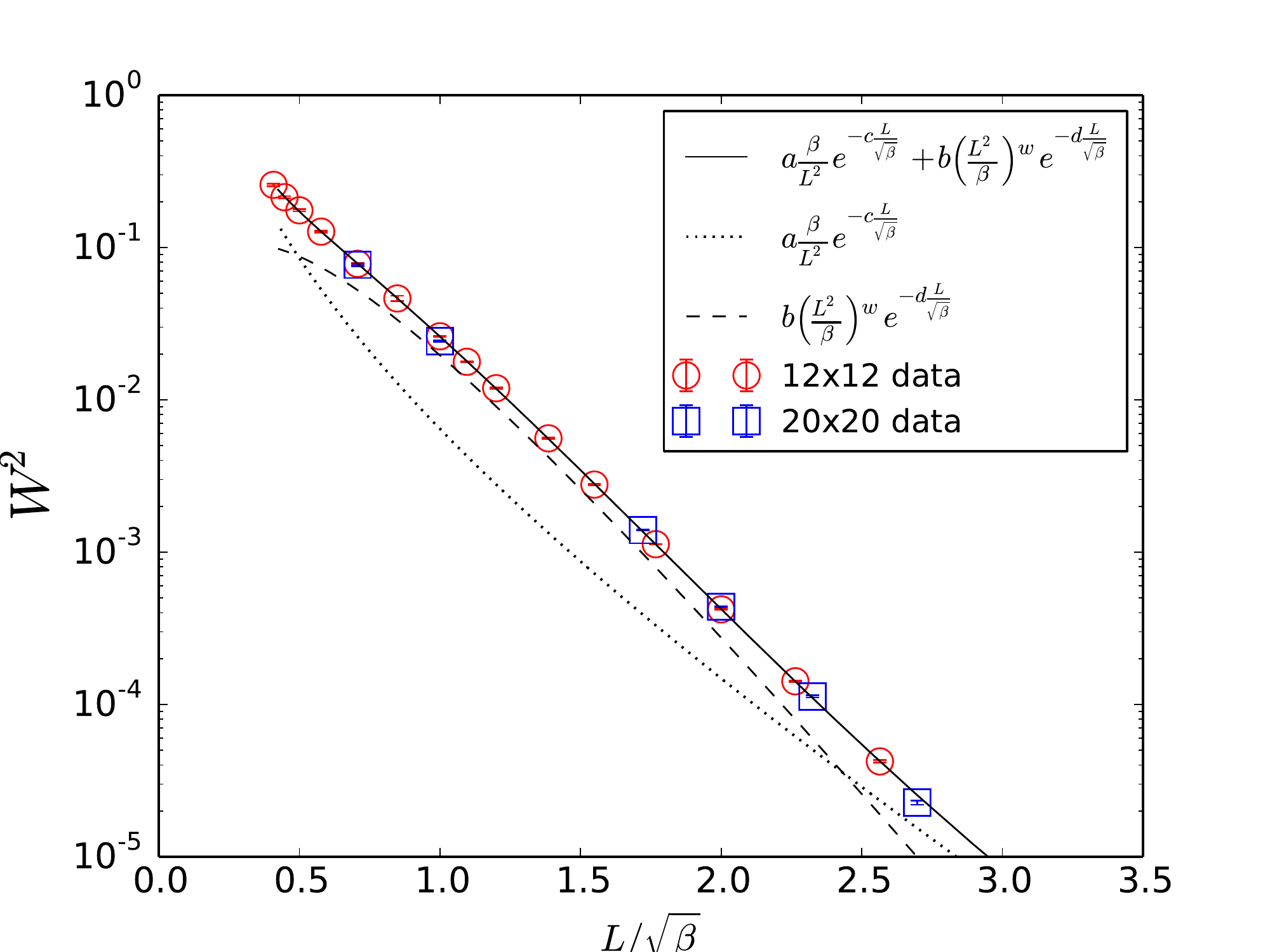}
\caption{
\label{fig:QRBetaStiff12}
(Color online) $\beta\rho$ as a function of $L/\sqrt{\beta}$ for the QR model. Results are shown for $L=12,20$.
The solid black line indicates a fit to the $L=12$ data of the form $a\frac{\beta}{L^2} e^{-c \frac{L^2}{\beta}}  + b \left(\frac{L^2}{\beta}\right)^{w} e^{-d\frac{L}{\sqrt{\beta}}}$
yielding \QRwtwelve.
The dotted line indicates the first part of this fit while the dashed line shows the second part.
}
\end{figure}

We now turn to the QR model in the presence of disorder. In this case we assume that the scaling variable
$L/\beta$ generalizes to $L/\beta^{1/z}$. We therefore expect to find:
\begin{equation}
W^2= a\frac{\beta}{L^z} e^{-c L/\beta^{1/z}}+b\left(\frac{L^z}{\beta}\right)^w e^{-d L/\beta^{1/z}}.
\end{equation}
In Fig.~\ref{fig:QRBetaStiff12} results are shown for $W^2$ for two different lattice sizes $L=12,20$
Assuming $z=2$ in this case we plot the results against $L/\sqrt{\beta}$ demonstrating that the results fall on a single
curve with only slight deviations from a straight line. It is perhaps surprising that it is the variable $L/\beta^{1/z}$ that
appears as opposed to $L^z/\beta$ but this can very clearly be verified from the simulations.
Performing a fit to the ansatz we find exceedingly good agreement with the expected form with a correction
exponent \QRwtwelve, close to the value for the model without disorder. An inspection of our results for $\rho$ for this system size shows that in this
case there is {\it no} maximum in $\rho$ versus $\beta$.

We have performed a similar analysis of $W^2$ as a function of $\beta$ for the QR model again clearly
confirming the overall exponential dependence and the presence of the two terms.

\section{Crossing of $\beta\rho$ for the QR model.}
\begin{figure}
\includegraphics[width=\linewidth]{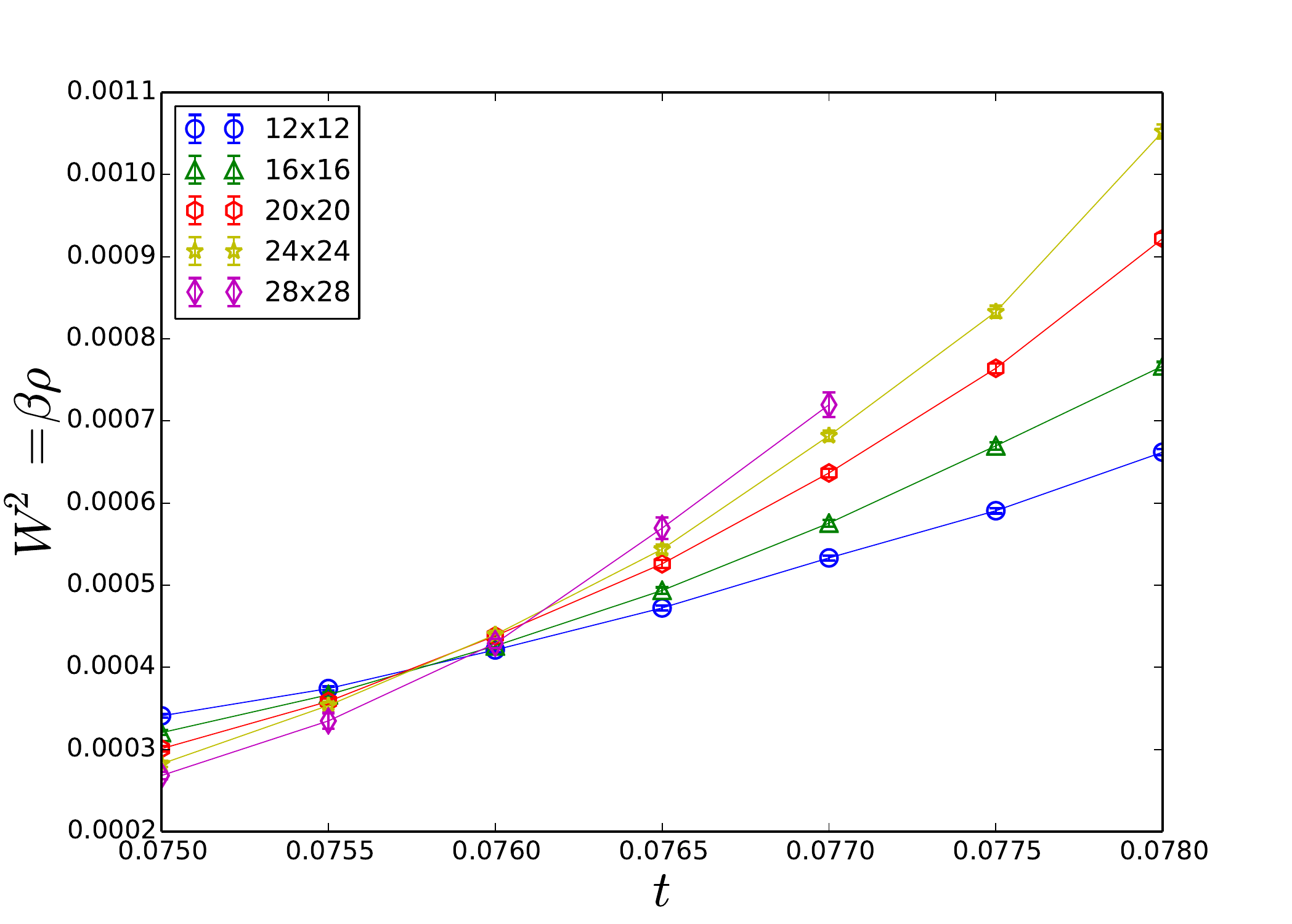}
\caption{
\label{fig:QRCrossing}
(Color online) $\beta\rho$ as a function of $t$ for the QR model. All simulations are performed using
the fixed aspect ratio $\beta=L^z/4$ with $z=2$. Lines cross at the critical point $t_c=0.0760(5)$.
}
\end{figure}

With the $z$ and $t_c$ determined from the fit in Fig.~\ref{fig:QRCollapse} $\beta\rho$ plotted for different
$L$ at a fixed aspect ratio $\beta=L^z/4$ should also cross at the critical point \QRtc. This is indeed the case and is
shown Fig.~\ref{fig:QRCrossing}. Since this data is essentially already shown in Fig~\ref{fig:QRCollapse}(a) we have in 
the main text opted to show the crossing using the related quantity $B_{W^2}=[\left<W^4\right>]/[\left<W^2\right>]^2$
shown in Fig.~\ref{fig:QRBinder}.

\end{document}